\documentclass[review]{elsarticle}

\usepackage{graphicx}
\usepackage{epstopdf}
\usepackage{color}

\usepackage{lineno,hyperref}
\modulolinenumbers[5]

\usepackage{xfrac}
\usepackage{fixltx2e}
\usepackage{epstopdf}
\usepackage{balance}
\usepackage{amsmath}
\usepackage{amsthm}
\usepackage{amsfonts}
\usepackage{amssymb}
\usepackage{makeidx}  % allows for indexgeneration
\usepackage{algorithm}
\usepackage{algpseudocode}
\usepackage{color}

\journal{Journal of Computers \& Electrical Engineering}

\begin{document}
	
\begin{frontmatter}
	
%
%\mainmatter              % start of the contribution
%

\title{Characterising Resource Management Performance in Kubernetes}

%
%\titlerunning{Kubernetes Performance}  % abbreviated title (for running head)

\author[zaragozaadd]{V\'{i}ctor Medel}
\ead{vmedel@unizar.es}

\author[zaragozaadd]{Rafael Tolosana-Calasanz}
\ead{rafaelt@unizar.es}

\author[zaragozaadd]{Jos\'{e} \'{A}ngel Ba\~{n}ares}
\ead{banares@unizar.es}

\author[zaragozaadd]{Unai Arronategui}
\ead{unai@unizar.es}

\author[cardiffadd]{Omer Rana}
\ead{ranaof@cardiff.ac.uk}

\address[zaragozaadd]{Aragon Institute of Engineering Research, University of Zaragoza, Spain}
\address[cardiffadd]{School of Computer Science \& Informatics, Cardiff University, UK}

\cortext[cor1]{Corresponding authors}

%\author{\IEEEauthorblockN{V\'{i}ctor Medel\IEEEauthorrefmark{1}, Omer Rana\IEEEauthorrefmark{2}, Jos\'{e} \'{A}ngel Ba\~{n}ares\IEEEauthorrefmark{1}, Unai Arronategui\IEEEauthorrefmark{1}}
%	\IEEEauthorblockA{\IEEEauthorrefmark{1}Aragon Institute of Engineering Research (I3A), University of Zaragoza, Spain \\
%		Email: vmedel@unizar.es, banares@unizar.es, unai@unizar.es}
%	\IEEEauthorblockA{\IEEEauthorrefmark{2}School of Computer Science \& Informatics, Cardiff University, UK \\
%		Email: ranaof@cardiff.ac.uk}}

%\authorrunning{Medel et al.}

% the affiliations are given next

%\toctitle{Lecture Notes in Computer Science}
%\tocauthor{Authors' Instructions}
%\maketitle

\begin{abstract}        % give a summary of your paper	

One of the challenges for enabling elastic automated resource management in cloud computing is to accomplish effective automated resource management actions, which include provisioning, maintaining, and de-provisioning of computing power. Among the cloud resources currently available, containers are rapidly replacing Virtual Machines (VMs) as the compute instance of choice in cloud-based deployments. One of the reasons is the significantly lower overhead of deploying and terminating containers in comparison to VMs. Understanding performance associated with deploying, terminating and maintaining a container is therefore significant. In this paper, we analyse performance of the Kubernetes system and develop a Petri net-based model of resource management within this system. Our model is characterised using real data from a Kubernetes deployment, and can be used as a basis to design scaleable applications that make use of Kubernetes.  

\end{abstract}


\end{frontmatter}

\section{Introduction}
\label{sec:intro}
%Omer

%\begin{itemize}
%\item
%Manage container deployment and services -- using a Reference net/Petri net based model
%
%\item
%Characterise the model with performance data from physical deployment
%
%\item
%Algorithm adaptation with resource properties  -- job types depending on
%capability available within resources
%
%\end{itemize}
%
%References:
%\url{http://wapco.e-ce.uth.gr/program.html}
%

Cloud systems enable computational resources to be acquired (and released) on-demand and in accordance with (changing) application requirements. Users can rent computational resources of different types: virtual machines (VMs), containers, specialist hardware (e.g. GPU or FPGA), or bare-metal resources, each having their own characteristics and cost. An effective automated control of cloud resource (de-) provisioning needs to consider~\cite{DBLP:conf/ucc/Tolosana-Calasanz16}: (i) resource utilization, (ii) economic cost of provisioning and management, and (iii) the resource management actions that can be automated. 
%Automation of management actions, such as resource (de-) allocation and configuration needs to be effective and adjust the computational power required to the actual Service Level Agreements (SLAs).
%There is a number of cloud providers that support resource provisioning on a per second or even millisecond basis (rather than on a per hour as it was in the past), such as AWS EC2~\footnote{https://aws.amazon.com/}, GCE~\footnote{https://cloud.google.com/}, or Amazon Lambda.~\footnote{https://aws.amazon.com/lambda/}
Increasingly, many cloud providers support resource provisioning (and billing) on a per second or even per millisecond basis, such as GCE\footnote{\url{https://cloud.google.com/}}, or Amazon Lambda\footnote{\url{https://aws.amazon.com/lambda/}} -- referred to as ``serverless computing''. 
%A lambda function is provisioned through a container-based deployment, whose execution is billed at 100ms intervals.
Therefore, understanding performance associated with deploying, terminating and maintaining a container that hosts that function is significant, as it affects the ability of a provider to offer finer grained charging options for users with stream analytics/ processing application requirements.
Provisioning and de-provisioning actions are subject to  a number of factors~\cite{DBLP:conf/ucc/Tolosana-Calasanz16}, mainly: (i) the \emph{overheads} associated with the action (e.g. launching a new VM can often take minutes~\cite{Ostermann2010}); and (ii) the actual processing time required can vary due to resource contention -- leading to uncertainty for the user. 

%Provisioning and de-provisioning actions in this context are subject to \emph{inertia}. From a control perspective, intuitively, inertia
%is the tendency of the system to remain unchanged for a time interval after taking an action.
%Such an inertia can be due to a number of factors~\cite{DBLP:conf/ucc/Tolosana-Calasanz16}, mainly: (i) the \emph{overheads} associated with the action (e.g. launching a new Virtual Machine (VM) can often take minutes~\cite{Ostermann2010}); (ii) the actual processing time required can vary due to resource contention -- leading to uncertainty for the user. 
%Such time delays often lead to uncertainty time intervals where the controller ignores whether the action has been ineffective or its effect still has not taken place. 

%Furthermore, they can even have worse undesired effects leading the system to instability, that is, alternately switching between increasing and decreasing computational power too fast and frequently, potentially causing potentially causing SLA violations.

Kubernetes~\cite{Burns:2016} is a system that enables a container-based deployment within Platform-as-a-Service (PaaS) clouds, focusing specifically on cluster-based systems. 
It can provide a cloud-native application (CNA)~\cite{kratzke2017understanding}, a distributed and horizontally scalable system composed of (micro)services, with operational capabilities such as resilience and elasticity support.
From an architectural point of view, Kubernetes introduces the \emph{pod} concept, a group of one or more containers (e.g. Docker, or any OCI compliant container system) with shared storage and network. 
%Kubernetes also enables deployment of multiple ``pods'' across physical machines, enabling scale out of an application with a dynamically changing workload.
%Each pod can allocate multiple containers, which can make use of services (e.g. file system and I/O) associated with a pod. Any OCI compliant container runtime engine could be used, but we chose Docker as it is the most popular engine for Kubernetes. 
In this paper, we investigate deploying, terminating and maintaining performance of (Docker) containers with Kubernetes, identifying operational states that arise with the associated \emph{pod}--container. This is achieved through Reference Nets (a kind of Petri-Net (PN)~\cite{murata1989petri}) based models. The models can be further annotated and configured with deterministic time, probability distributions, or functions obtained from monitoring data acquired from a Kubernetes deployment. It can also be used by an application developer / designer: (i) to evaluate how pods and containers could impact their application performance; or (ii) to support capacity planning for application scale-up / scale-down.

This paper extends~\cite{Medel:2016bb} by: (i) the inclusion of additional experiments in a larger cluster; (ii) considering the impact of variable latency/Round-Trip Time (RTT) in the communication network; (iii) analysing the impact of varying the number of containers inside a pod; (iv) analysing the impact of downloading a container image at deployment time; (v) using rules to assist developers to better structure their Kubernetes deployment. The paper is organized as follows. In Section~\ref{sec:state}, we describe our model. Section~\ref{sec:experiment} shows our pod abstraction overhead characterization. We discuss the deployment results in Section~\ref{sec:discus} and related work in Section~\ref{sec:related}. The conclusions are  outlined in Section~\ref{sec:conc}. % We present a brief description on Reference Nets and Kubernetes in Appendices A, B. %\ref{sec:background} and \ref{sec:dockerkub}.

\section{Kubernetes Overhead Analysis \& Performance Models}
\label{sec:state}
The Kubernetes architecture incorporates the concept of a pod, an abstraction that aggregates a set of containers with some shared resources at the same host machine. It plays a \emph{key} factor in the overall performance of Kubernetes.% and we will consider it in this section.
We make use of Reference Nets to model pods and containers and to conduct performance analysis.  Reference Nets can be interpreted by Renew\cite{kummer2004extensible}, a Java-based Reference net interpreter and a graphical modelling tool.

\subsection{Kubernetes Performance Model}
\label{sec:characperf}

Kubernetes supports two kinds of pods: 
(i)\emph{Service Pods}: They are run permanently, and can be seen as a background workload in the cluster. Two key performance metrics are
associated with them: (i.a) availability (influenced by faults and the time to restart a pod/container) and (i.b) utilisation of the service (impacting response time to clients).  For example, high utilisation leads to an increased response time. Several Kubernetes system services (e.g. container network or DNS) and high level services (e.g. monitoring, logging tools) are provided by Service Pods.  
(ii)\emph{Job/batch Pods}: They are containers that execute tasks and terminate on task completion. For a Job pod, both deployment and total execution time (including restarting, if necessary) are important metrics. The restart policy of these containers can be \textit{onFailure} or \textit{never}.

When a pod is launched in Kubernetes, it requests resources (RAM and CPU) to the Kubernetes scheduler. If enough resources are available, the scheduler selects the {\it best} node for deployment. The requested CPU could be considered as a reservation in contingency situations. For instance, when a container is idle (e.g. it is inside a service pod and the service has low utilisation), other containers can use the CPU.  With this resource model, the overall performance of the pod depends on its resource requests and on the overall workload.  We could define a CPU usage limit, but then some resources might remain unused.

\begin{figure}[t]
	\centering
	\includegraphics[width=0.95\linewidth]{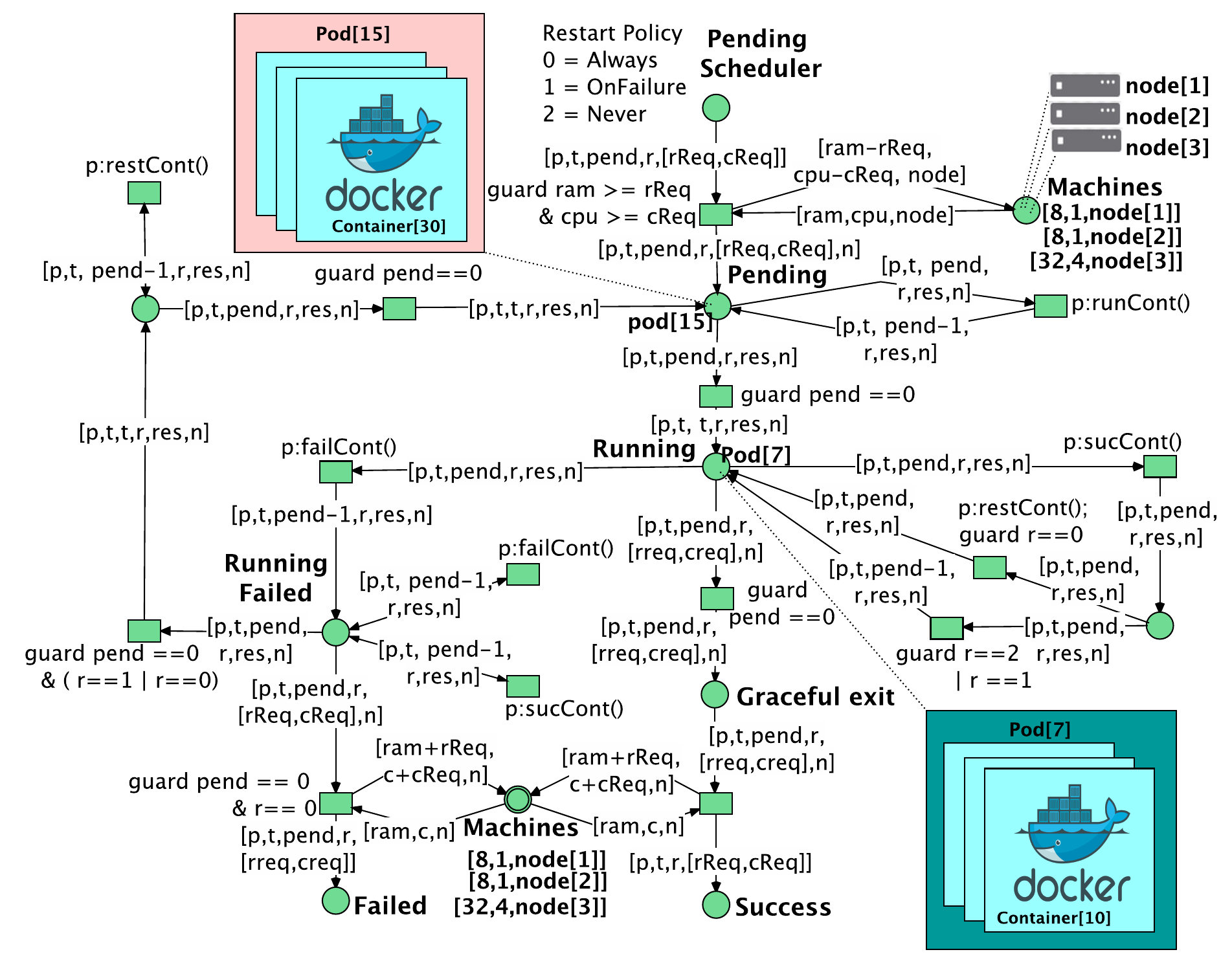}
	\caption{Model of the life cycle of  pods in Kubernetes.}
	\label{fig:PodLifecycle}
\end{figure}

\begin{figure}[t]
	\centering
	\includegraphics[width=0.7\linewidth]{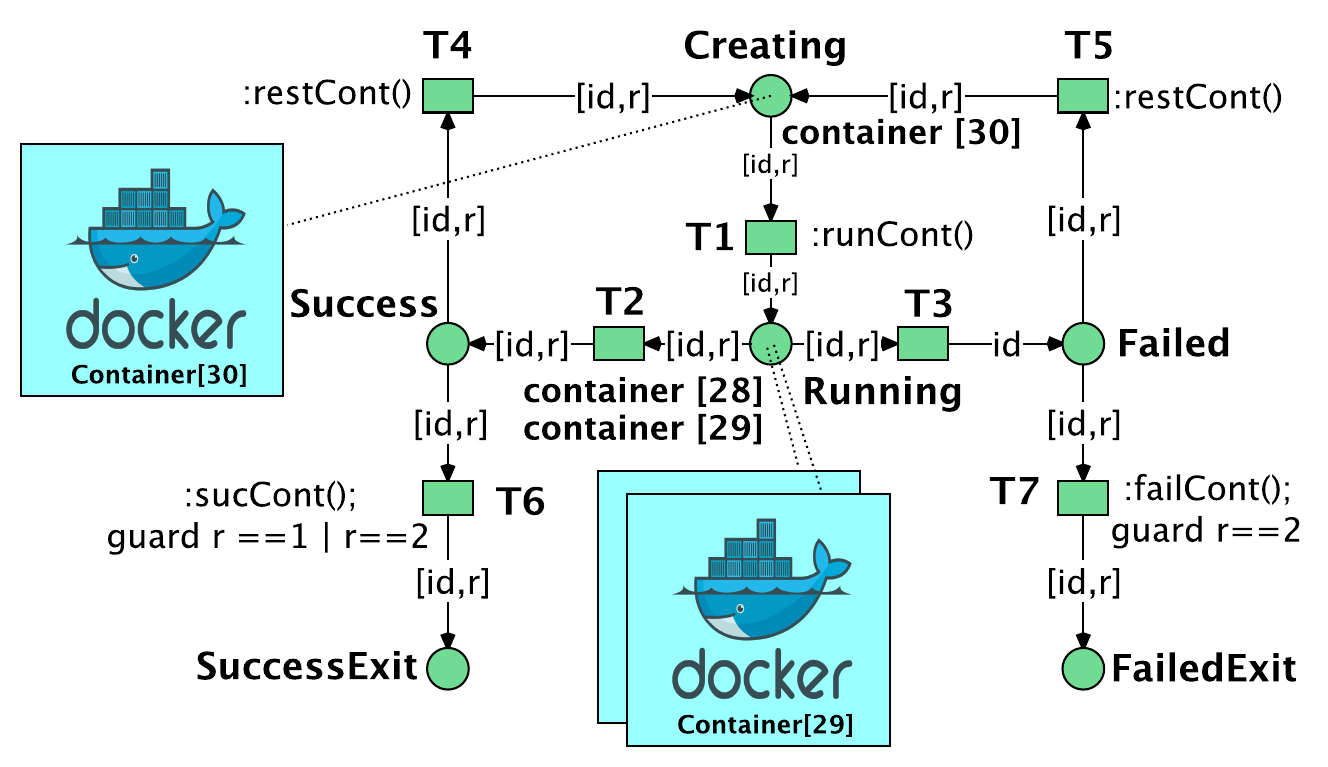}
	\caption{ Model of the life cycle of  containers inside a pod. $r$ models the restart policy of a container -- Always = 0, OnFailure=1, Never = 2.}
	\label{fig:ContainerLifecycle}
\end{figure}

We model a pod's life cycle in order to estimate the impact of different scenarios on the deployment time and the performance of the applications running inside a pod. In Kubernetes, a pod's life cycle depends on the state of the containers that are inside it.  For instance, a pod has to wait until all its containers are created. 
With the Reference Nets abstraction, we can provide an unambiguous hierarchical representation of the Kubernetes manager system as the System Net and the Pods (with the containers) as the Token Nets. The tokens inside our Token Net represent containers and the tokens inside our System Net represent Pods, as illustrated in Figures~\ref{fig:PodLifecycle} and~\ref{fig:ContainerLifecycle}. The models were derived from the Kubernetes documentation~\footnote{\url{https://kubernetes.io/docs}}; specifically, from the Pod Lifecycle section~\footnote{\url{https://kubernetes.io/docs/concepts/workloads/pods/pod-lifecycle/}} and from the Resource Management section~\footnote{\url{https://kubernetes.io/docs/concepts/configuration/manage-compute-resources-container/}}. Details  about  places and transitions, needed to specify the initial marking,  are hidden to improve legibility.  In addition, we assume that the scheduler assigns a pod to a single node arbitrarily, as long as the machine has enough resources available. %(A more appropriated representation is the representation of a Kubernetes cluster as the system-net, which manage the pod's life cycle and perform the mapping of pods to resources. For the sake of simplicity, we have include part of this functionality in the pod model, removing the cluster level.)
If there are not enough resources in the cluster, the pod waits in \textbf{Pending Scheduling} place. This behaviour could be refined by introducing more sophisticated policies and a rejection place for pods. \textbf{Machines} place~\footnote{It should be noted that \textbf{Machines} place appears twice: One with a single circle (actual definition) and with a double circle (a duplication to simplify the model). Reference nets support double circle to simplify the model and to improve its legibility.  If it were not used, several arcs would cross the model with their corresponding arc labels} represents the resources managed by the scheduler. For each machine, there is a tuple token with the identification of the node, the available RAM size and number of available cores. Figure~\ref{fig:PodLifecycle} shows three machines ranging from 8GB to 32GB,  with 1 to 4 cores.  The resources assigned to a pod are only released when the pod restart policy is ``never" or ``onFailure".

Once the pod has been assigned to a machine, Kubernetes starts creating the containers  -- it is in \textbf{Pending Scheduler} place -- while the pod waits in its \textbf{Pending} place.  Both nets are synchronized through the inscription  \textit{runCont}. In this way, when  a  container in a  Pod  changes to \textbf{Running} place in Figure~\ref{fig:ContainerLifecycle}, the number of pending containers in this pod  is decremented in the  \textbf{Pending} place of Figure~\ref{fig:PodLifecycle}. When all containers are running in the pod, the transition with the  \textbf{guard pend==0}  is fired and the pod states changes to \textbf{Running}.

%Once the pod has been assigned to a machine, Kubernetes starts creating the containers  -- it is in \textbf{Pending Scheduler} place -- while the pod waits in its \textbf{Pending} place.  The System Net model -- Figure~\ref{fig:PodLifecycle} -- synchronizes with the  Token net -- Figure~\ref{fig:ContainerLifecycle} -- through the inscription  \textit{runCont}. In this way, when  a  container in a  Pod  changes to state  \textbf{Running} in Figure~\ref{fig:ContainerLifecycle}, the number of pending containers in this pod  is decremented in the  \textbf{Pending} place of Figure~\ref{fig:PodLifecycle}. When all containers are running in the pod, the transition with the  \textbf{guard pend ==0}  changes  the pod state to  \textbf{Running}. 

While the pod is in \textbf{Running} state, it is waiting for its containers to terminate. If a container fails, the pod goes to \textbf{RunningFailed} place where it waits for the termination of all containers (with a potential restart action).  If there are no failures, the pod will be in \textbf{Running} place or eventually will reach \textbf{Success} place when all containers have finished. 

Figure~\ref{fig:ContainerLifecycle} illustrates the behaviour of a container. A token in that net represent a container. A pod's restart policy is included in the net. A created pod enters the \textbf{Running} place, and may reach the \textbf{Success} or \textbf{Failure} place.  The firing of the corresponding Transitions (\textbf{T1}, \textbf{T2} and \textbf{T3}) is synchronised with the System Net.  According to the restart policy, the containers might return to \textbf{Running} place or they might finish in \textbf{SuccessExit} or in \textbf{FailedExit} places. We include several timed transitions, as summarised in Table~\ref{tab:timedTransitions}.  By default, the firing of \textbf{T2} and \textbf{T3} is arbitrary and non-deterministic; however, with Renew, it is possible to simulate any probability distribution for them in order to simulate a failure.  Additionally, it is possible to assign different random distributions for the timed transitions. In the next sections, we describe different experiments to obtain the real value of these metrics.  The termination time (\textbf{T6} and \textbf{T7}) and the termination time when a container is restarted (\textbf{T4} and \textbf{T5}) do not depend on the success of the container, so both transitions are modelled with the same distribution. When a container is restarted, the total restarting time can be calculated as \textbf{T4} -- or \textbf{T5}-- + \textbf{T1}.
%, so they are measured in Section \ref{sec:analysis}.

\newcommand{\splitbox}[1]{%
  \linespread{1}\selectfont
  \renewcommand{\arraystretch}{0.9}%
  \begin{tabular}{@{}l@{}}
    \strut#1\strut
  \end{tabular}%
}

%\footnotesize
\begin{table}
\footnotesize
	\renewcommand{\arraystretch}{0.9}
	\centering
	\small
	\begin{tabular}{| c |l| c | l |}
		\hline
		\textbf{Transition} & \textbf{Variable} & \textbf{Trans.} & \textbf{Variable}  \\
		\hline
		\textbf{T1} &  Container creation & \textbf{T2} & Container execution \\
		\textbf{T3} & Time to Container failure &  \textbf{T4}, \textbf{T5} & \splitbox{Container termination time \\ if the container is restarted } \\
		\textbf{T6}, \textbf{T7} &  \splitbox{(Graceful) \\ Container termination} &  & \\
		\hline
	\end{tabular}
	\caption{Timed transitions in the model}
	\label{tab:timedTransitions}
\end{table}
\normalsize

%\small
%\begin{table}
%	\centering
%	\caption{Timed transitions in the model}
%	\label{tab:timedTransitions}
%	\begin{tabular}{| c | c |}
%		\hline
%		Transition & Variable \\
%   		\hline
%		T1 & Time to create a container \\
%		T2 & Execution time of a container \\
%		T3 & Time until next failure in a container \\
%		T4,T5 & Time to restart a container. \\
%		T6, T7 & Time to finish gracefully a container. \\
%		\hline
%	\end{tabular}
%\end{table}
%\normalsize

\subsection{Experiments to feed the performance model}
We conducted several experiments to estimate the value of transitions in Table~\ref{tab:timedTransitions} by deploying Kubernetes on a cluster with eight physical machines --$n=8$--, each with 32GB of RAM and 4 Intel i5-4690(3.500GHz) cores. The results are shown in the next subsections.  The performance metrics of the high level model (Figure~\ref{fig:PodLifecycle}) are determined by the firing sequences of transitions in the Token Net (Figure~\ref{fig:ContainerLifecycle}).  For example, if there is a pod with three containers, the \textbf{T1} transition of this pod is fired three times.  The pod is waiting this time in the \textbf{Pending} place.

\subsubsection{Benchmarking Starting Time}
\label{sec:creationtime}

%We conducted an experiment to estimate the value of Transition \textbf{T1} by deploying Kubernetes on a cluster with eight physical machines, each with 32GB of RAM and 4 Intel i5-4690(3.500GHz) cores.
To estimate the value of Transition \textbf{T1}, we launched a variable number of containers, whose image was preloaded at all the machines, and measured the total deployment time. Each experiment was repeated 30 times, and we calculated the mean, the standard deviation, and the confidence intervals (for $\alpha=0.05$).  In order to calculate the confidence interval, we assumed (by the Central Limit Theorem) that the underlying distribution of the sampled mean follows a normal distribution.

\begin{figure}[t!]
	\centering
	\makebox[\columnwidth]{\includegraphics[width=\textwidth]{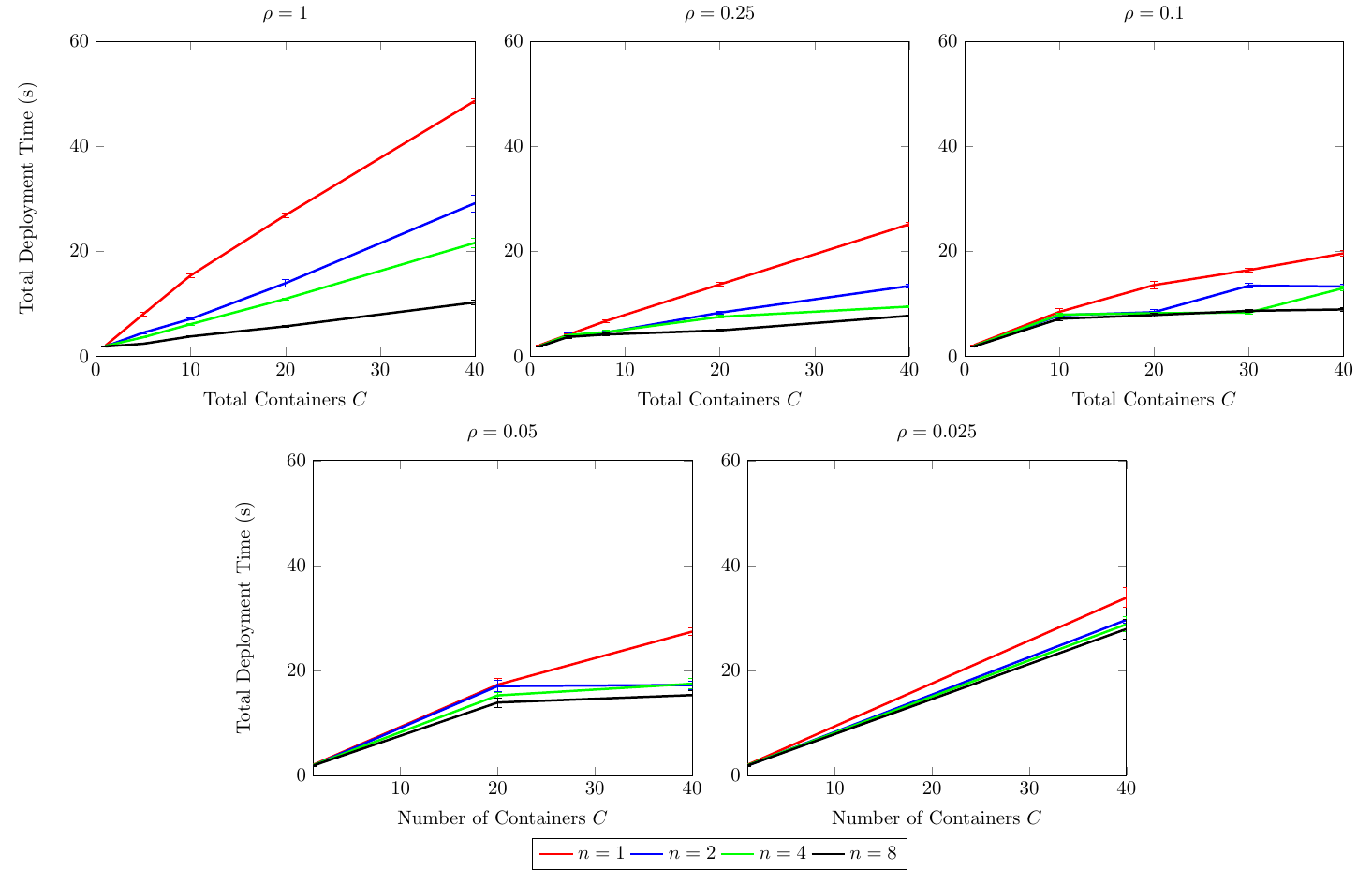}}
	\caption{Total deployment time ($T_d$) vs. Number of deployed containers ($C$).  Each graph shows: mean time, confidence interval for the mean for a varying number of machines in cluster, $n$.  The results are grouped by the number of containers inside a pod, $\frac{1}{\rho}$}
	\label{graph:deployment}
\end{figure}

In Figure~\ref{graph:deployment}, we show how different variables influence the deployment time. These variables are:
(i) The \emph{number of machines} available in the cluster ($n$).  We observe that the scheduler launches pods sequentially, without multi-threading (red line in Figure~\ref{graph:deployment}), showing a linear total deployment time with the number of deployed containers.
(ii) The \emph{number of containers $C$ inside a pod} ($\rho$ factor).  The $\rho$ factor is calculated as follows: $\rho=\frac{\#Pods}{C}$.  
For instance, a  $\rho$ factor of 0.25 implies that there are 4 containers inside each pod.  We can see that the time to deploy 10 pods with 4 containers in a single machine is 25.89 s., which is higher than the time to deploy 10 pods in a single container on a single machine (16.01 s.).
(iii) \emph{Cluster constraints} as the number of nodes.  

\begin{figure}[t!]
	\centering
\makebox[\columnwidth]{\includegraphics[width=\textwidth]{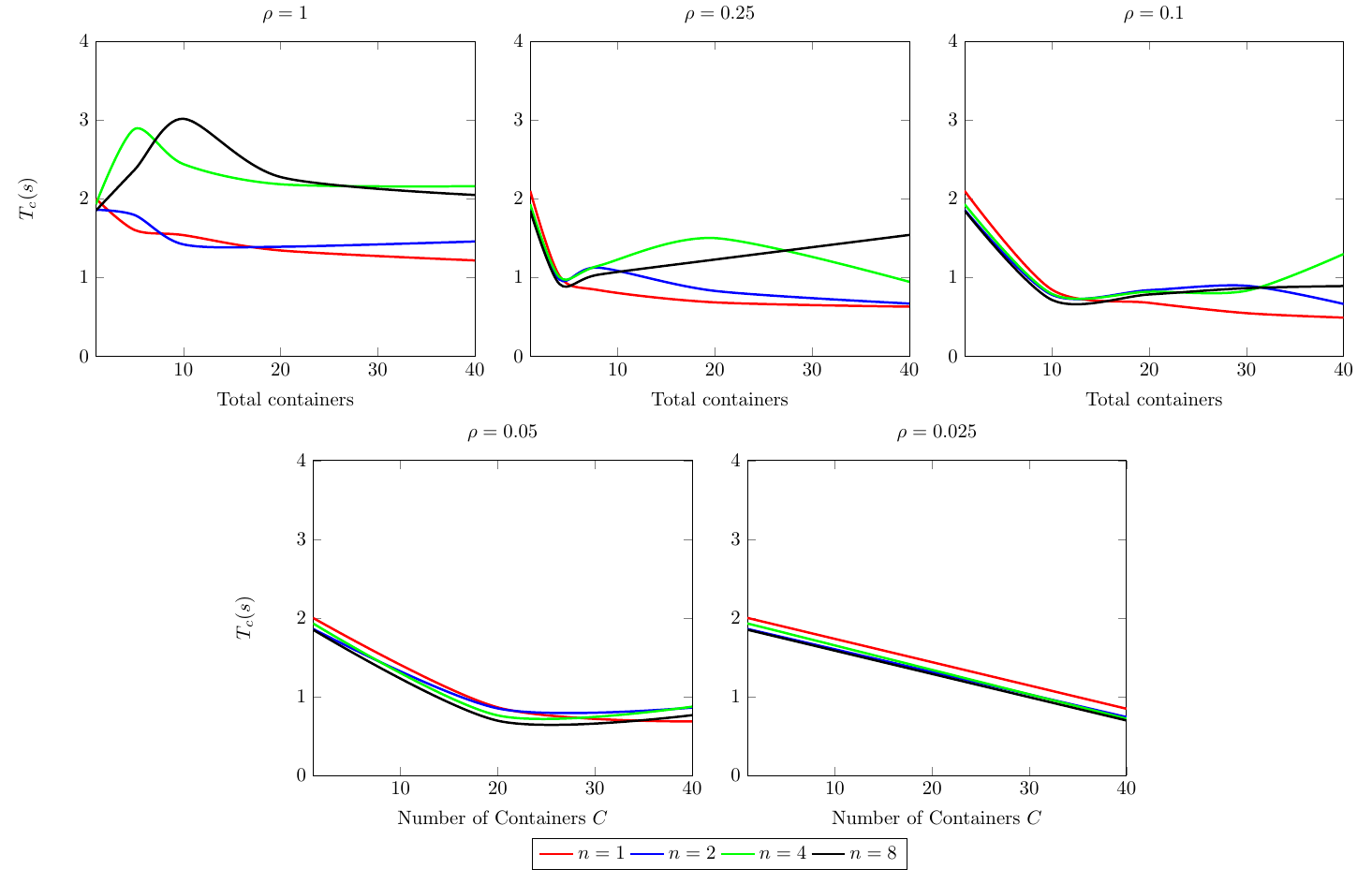}}
	\caption{Time to create a \emph{single} container ($T_c$ function) vs. Number of deployed containers ($C$).   Each graph shows: mean time, confidence interval for the mean for a varying number of machines in cluster, $n$.  The results are grouped by the number of containers inside a pod, $\frac{1}{\rho}$}
	\label{graph:deployment2}
\end{figure}

The total provisioning time ($T_t$) is calculated as: $T_t=T_d+T_{down}$
Where $T_d$ is the time to deploy pods and containers on physical machines and $T_{down}$ is the time to download the needed container image to the involved machines. Our experiments show that $T_d$ has a linear behaviour. Therefore, $T_d$ depends on the number of deployed containers $C$ and on the number of deployed pods, $\#Pods$: $T_d=\frac{C \ T_c(\rho, n, C)}{\min\{\#Pods, n\}}$.  As the scheduler manages the pod as the minimal schedulable unit, the maximum number of pods deployed in parallel in a cluster is given by $\min\{\#Pods, n\}$. $T_c$ is a function that returns the time to create a single container.  This value depends on how the deployment is structured -- $\rho$ and $C$ parameters -- and the number of machines in the cluster ($n$).  %We observe that $T_c$ changes when we decrease $\rho$ or increase the number of machines in the cluster.  
In Figure~\ref{graph:deployment2}, we show different values for $T_c$, obtained experimentally. We can see that for large $C$ values, and as $\rho$ approaches 0,  $T_c$ becomes constant.  Therefore, we can write:
$\lim_{\rho\to0, n\to\infty, C\to\infty} T_c(\rho, n, C) = t_c$.  Under these assumptions, the $T_c$ value could be considered as a constant -- attached to \textbf{T1} transition.

\begin{figure}[t!]
	\centering
\makebox[\columnwidth]{\includegraphics[width=\textwidth]{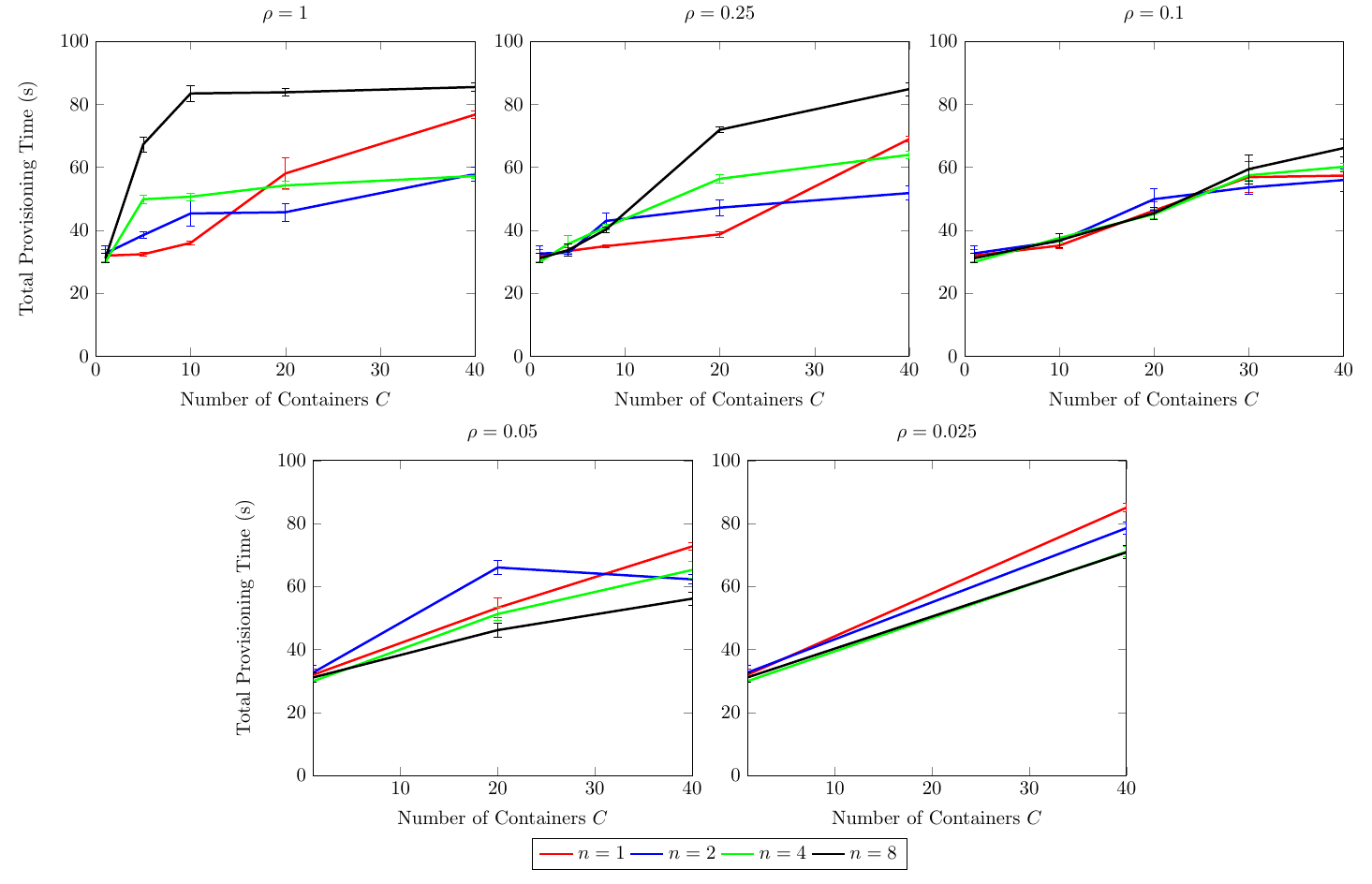}}
	\caption{$T_t$ vs. $C$.   Each graph shows: mean time, confidence interval for the mean for a varying number of machines in cluster, $n$.  The results are grouped by the number of containers inside a pod, $\frac{1}{\rho}$. The container image (1.225 GB) is not present in the machines.}
	\label{graph:deployment3}
\end{figure}

In order to characterise the impact of container image download time, we repeated the previous experiments without preloading the image. The image is downloaded from a machine located inside the cluster connected directly to the same switch.  The results are shown in Figure~\ref{graph:deployment3}. The results of the homogeneous latency scenario are better than the ones of the variable latency scenario -- as $\rho$ tends to zero, the latency impact on the results decreases.  We can see that as the number of machines increase, the total deployment time also increases, because the image needs to be downloaded by all the machines in the cluster, and all the machines are connected to the same server.  When the number of deployed pods is greater than the number of machines, the deployment time remains stable, so we can conclude that Kubernetes only downloads the image once per machine.  

%We deploy Kubernetes on two physical machines, each with 32GB of RAM and 12 Intel Xeon E5-2620 (2.00GHz) cores.  In Figure\ref{graph:deployment}, we show the mean deployment time of the Guestbook application with different configurations with one and two machines (n=1, n=2). 
%Each machine has pre-loaded docker images.  In Guestbook experiments, each pod has exactly one container.  The ``jobs" line in the figure is the same experiment with all containers in a single pod.  From the result, we can observe the overhead introduced by creating pods over the docker containers.  Figure~\ref{graph:deployment2} depicts the mean time needed to create a container per machine in each scenario, illustrating $\sim$ 0,6s deployment time for a single machine and 1s for two machine. A possible explanation of this behaviour could be the overhead introduced to synchronize the deployment in different machines.

%\begin{figure}[t]
%	\centering
%	\includegraphics[width=0.3\textwidth]{figures/graph2.pdf}
%	\caption{Mean Deployment time per container per machine vs. number of containers.}
%	\label{graph:deployment2}
%\end{figure}

%  \begin{figure}[!t]
%  	\subfloat[Deployment time vs. Number of Pods with a 95\% confidence interval.]{%
%  		\includegraphics[width=0.3\textwidth]{figures/graph1.pdf}\label{graph:deployment}
%  	}
%  	\hfill
%  	\subfloat[Mean Deployment time per container per machine vs. Number of Pods.\label{graph:deployment2}]{%
%  		\includegraphics[width=0.3\textwidth]{figures/graph2.pdf}
%  	}
%  	\caption{Comparing deployment time}
%  	\label{fig:Deployment time}
%  \end{figure}

Several variables related to the cluster architecture impact the deployment time, such as parameters of the physical machines and the network topology.  To assess the network topology impact, we repeated the experiments in a cluster with heterogeneous latency.  We simulated that half of the machines are in a different network area, so that their Round Trip Time (RTT) is about 100ms. The RTT for the rest of the machines is 0.25ms. Table~\ref{tab:latency} depicts the results for $\rho=1$ and $n=8$.  The results for other values of $\rho$ and $n$ are quite similar.  We can see that the latency has not a significant impact on $T_d$ -- and neither on $T_c$.  As in $T_t$ is included the time to download the container image, this value is higher.  However, the size of the image mitigates the latency impact.

\begin{table}[t]
\footnotesize
    \renewcommand{\arraystretch}{0.9}
	\centering
	\begin{tabular}{| c | c | c | c | c|}
		%	\hline
		\multicolumn{1}{c}{}& \multicolumn{1}{c}{\textbf{Homogeneous RTT}} & \multicolumn{1}{c}{} & \multicolumn{1}{c}{\textbf{Heterogeneous RTT}} & \multicolumn{1}{c}{} \\
		\hline
		\boldmath$C$ & \boldmath$T_d$ & \boldmath$T_t$ & \boldmath$T_d$ & \boldmath$T_t$ \\
		\hline
		1	& 1.85  &31.09	& 1.94 & 33.26 \\
		5	& 2.37	& 67.28 & 2.66	& 66.79 \\
		10	& 3.77	& 83.44  & 3.87	& 82.66  \\
		20  & 5.69	& 83.81 & 5.56	& 86.93 \\
		40 & 10.24 & 85.47 & 10.21 & 91.02 \\
		\hline
	\end{tabular}
	\caption{$T_d$ and $T_t$ values from a Kubernetes cluster with homogeneous RTT (0.25ms) and from a Kubernetes cluster with heterogeneous RTT. $\rho=1$ and $n=8$. The container image is 1.225 GB. The results are in seconds.} 	\label{tab:latency}
\end{table}

\subsubsection{Benchmarking Termination Time}

A Pod is expected to be terminated at some time.  If it is a service, and consequently it has to be running all the time, the termination may be due to a failure and the pod has to be restarted.  This philosophy also applied to containers, as we discussed previously in the Container Net model (transitions \textbf{T4}, \textbf{T5}, \textbf{T6}, \textbf{T7}).  We consider \textbf{T3} to depend on the application, and it represents the failure rate (or the time between failures). When a pod terminates, Kubernetes waits for a grace period (which, by default, is 30 seconds) until it kills any associated container and data structures.  

As far as we have tested, the only variable that affects termination time of a pod is the number of containers in that pod.  This occurs because when a pod finishes, all its containers have to be finished; however, in a normal scenario, pods finish -- or restart -- asynchronously.  Therefore, the overhead caused by finalising several pods on several machines is negligible. As $\rho$ approaches zero, the mean time to stop a single container inside a pod remains constant. The way in which these times are aggregated and synchronised depends on the scenario, and the specific performance metrics can be derived from the complete Petri Net Model.  

In order to associate the corresponding metric for the transitions, we perform the experiments shown in Table~\ref{tab:TerminationTime} in the same cluster, as in the previous section.  We present the results for \textbf{T4} and \textbf{T6} which correspond to a successful scenario.  Without taking into account the time to detect the failure, the behaviour of transitions \textbf{T5} and \textbf{T7} is similar: (i) \emph{Transitions \textbf{T4, T5}}: These transitions measure the time to stop a container when it is going to be restarted.  We have deployed pods with a variable number of containers  to measure the time. The results are shown in column ``\textbf{T4} per Container" in Table~\ref{tab:TerminationTime}.   When we decrease $\rho$,  the mean time to terminate a container remains constant. Additionally, the highest measured mean time is $\sim$ 0.3s  and 80\% of sampled times are $<$ 0.22s. (ii) \emph{Transitions \textbf{T6, T7}}: These transitions model the normal behaviour of Kubernetes.  On successful completion, Kubernetes waits for the grace period and deletes all the data structures associated with a container.  We measured these variables in columns ``\textbf{T6} Graceful termination" and ``\textbf{T6} per container" in Table \ref{tab:TerminationTime}.  For these experiments, we set the grace period to 30s (the default).  We can observe that the stopping time remains constant for more than 10 containers in a pod (Column ``\textbf{T6} per container") and for low values is negligible. It is interesting to note that the time to stop a container is higher when the container is going to be restarted.  This overhead is about 10ms.

%\footnotesize
\begin{table}
\footnotesize
	\renewcommand{\arraystretch}{1.1}

	\centering
	\begin{tabular}{| c |  c | c | c| c| }
		\hline
		\boldmath$C$ & \boldmath$\rho$ & \splitbox{\textbf{T4 (T5) per} \\ \textbf{Container}} & \splitbox{\textbf{T6 (T7) Graceful} \\ \textbf{termination}} &  \splitbox{\textbf{T6 (T7) per} \\ \textbf{Container}} \\
		\hline
		1 &	1 &	0.01  & 30	&  0 \\
		10 & 0.1 &	 0.11 & 30.99	&  	0.10 \\
		20 & 0.05 &	 0.12 & 32.8	    &  0.14 \\
		%30 &	4.133958016	& 0.137798601 & & \\
		40 & 0.025 &	0.15 & 34.69 &		0.11 \\
		% 50 &	7.12	& 0.14 & 35.85 &	5.85 &	0.11  \\
		60 & 0.016  &  0.16 & 37.04 &	 0.11 \\
		%70 &	25.77	& 0.37 & 37.56 &	7.56	& 0.10 \\
		%80 &	13.65396099 &	0.170674512 & &  \\		
		\hline
	\end{tabular}
		\caption{\textbf{T4} and \textbf{T6} experimental results (in seconds).}
	\label{tab:TerminationTime}
\end{table}
%\normalsize

%Commented before Cloud ACM changes.
%\footnotesize
%\begin{table}
%	\renewcommand{\arraystretch}{1.1}
%	\caption{T5 and T6 experimental results}
%	\label{tab:TerminationTime}
%	\centering
%	\begin{tabular}{| c | c | c | c| c| c| }
%		\hline
%		C & T5 per & T5 per & Graceful & T6 without & T6 per \\
%       \  & Pod & Container & termination & grace period & Container\\
%		\hline
%1 &	0.01 &	0.01  & 30	& 0	& 0 \\
%10 &	1.06	& 0.11 & 30.99	&  0.99 &	0.10 \\
%20 &	2.34	& 0.12 & 32.8	& 2.89	& 0.14 \\
%30 &	4.133958016	& 0.137798601 & & \\
%40 &	6.06	& 0.15 & 34.69 &	4.69 &	0.11 \\
% 50 &	7.12	& 0.14 & 35.85 &	5.85 &	0.11  \\
%60 &	9.61	& 0.16 & 37.04 &	7.04	& 0.11 \\
%70 &	25.77	& 0.37 & 37.56 &	7.56	& 0.10 \\
%80 &	13.65396099 &	0.170674512 & &  \\		
%		\hline
%	\end{tabular}
%\end{table}
%\normalsize

%Reference nets + use here

%\section{Scenarios \& Analysis}
%\label{sec:analysis}
%
%Omer to start ...
%
%
%Scenarios of use
%
%Increase variability in machines and access to the data -- introduce background workload into the cluster

%\section{Experimentation}
\section{Overhead Analysis of the Pod Abstraction}
\label{sec:experiment}
% Victor, Unai

%Experiments to measure the overhead of the pod abstraction and latency in pod deployments.

%\subsection{Measuring Overheads}

%o measure overheads for applications deployed in Kubernetes, we consider the following scenarios:
The pod abstraction allows several containers to be grouped together sharing different resources. However, the way in which resources are shared between containers in the same pod and the impact on the performance of a container are not easy to determine. In this section, we analyse this performance change based on how the deployment is structured (e.g. the number of containers inside each pod), using the total execution time as a metric. We performed several experiments to measure the overhead induced by the pod abstraction. The aim is to measure how transition \textbf{T2} is affected by the deployment configuration. 

Let us consider the following scenarios:
(i) \emph{Scenario 1}: A pod is deployed and all the containers are inside that pod -- $\rho =1/c$. 
(ii) \emph{Scenario 2}: Several pods are deployed and there is exactly one container inside each pod -- $\rho=1$.

The total number of containers deployed is given by $C$ and all pods are deployed on the same machine.  The machine has 12 Intel Xeon E5-2620 (2.00GHz) cores and 32GB of RAM.  Each experiment, one for each scenario, was repeated 30 times, so that we can consider that the probability distribution of both means follows a normal distribution (by the Central Limit Theorem).  We present the mean execution time ($\mu_i$) and the standard deviation ($\sigma_i$). In order to compare both scenarios, we propose the following statistical hypothesis test:
\begin{center}
$\left\{
\begin{array}{c}
H_0: \mu_1-\mu_2=0 \\
H_1: \mu_1-\mu_2 \neq 0 \\
\end{array}
\right.$
\end{center}

As we assume that both means follow a normal distribution and they have the same variance, we can use the Student t test~\cite{student1908probable}. Additionally, as there are several resources shared between containers, we can expect different behaviour for each one.  In the following subsections, we accomplished a hypothesis test for applications with high CPU usage -- Pov-Ray 3.7--, high I/O usage -- IOzone benchmark -- and high network usage --netperf.

\subsection{CPU intensive application}
%\subsubsection{CPU intensive application}

We used the multi-threaded pov-ray 3.7 application as a benchmark to measure the overhead of pods for CPU intensive use.  Kubernetes inherits from Docker the CPU quota reservation. This contingency mechanism allows a container to reserve a maximum CPU quota.  However, the quota is only used when there is contingency in the machine, otherwise, all available CPU is used.  The comparison between Scenarios 1 and 2 is presented in Table~\ref{tab:pov-ray}.  We can see that when the number of containers increases, the null hypothesis should be rejected.  Additionally, when $H_0$ is rejected,  Scenario 1 is faster than Scenario 2.  The overhead caused by having one container inside each pod is about 0.01\%.    % The results of the experiments are shown in Table \ref{tab:pov-ray}.  As expected, the execution time is linear to the number of parallel containers/pods executing pov-ray benchmark.  With a single container, all the CPU is used by the application.  As the machine has 12 cores, the performance of twelve containers should be similar to the performance of pov-ray executed on a Docker container with multi-threading disabled. With this metric as reference value, we can calculate the overhead introduced by Kubernetes in CPU usage (about 14\%).  With these results, it seems that, for CPU intensive applications, it is a better solution to group all containers in a same pod to reduce the overhead. \\

%\footnotesize
\begin{table}[t]
\footnotesize
    \renewcommand{\arraystretch}{0.9}
	\centering
	\begin{tabular}{| c | c | c | c | c|c|}
		%	\hline
		\multicolumn{1}{c}{}& \multicolumn{1}{c}{\textbf{Scenario 1}} & \multicolumn{1}{c}{} & \multicolumn{1}{c}{\textbf{Scenario 2}} & \multicolumn{1}{c}{} & \multicolumn{1}{c}{} \\
		\hline
		\boldmath$C$ & \boldmath$\mu_1$ & \boldmath$\sigma_1$ & \boldmath$\mu_2$ & \boldmath$\sigma_2$ & \boldmath$\mu_1 - \mu_2 = 0$? \\
		\hline
		1	& 123.47 &	0.43	& 123.38 &	0.39 & Yes \\
		4	& 473.65	& 0.96 & 475.15	& 0.62 & No \\
		8	& 946.90	& 0.72  & 946.63	& 0.69 & Yes  \\
		12	& 1417.76	& 1.67 & 1420.40	& 1.35 & No \\
		20	& 2370.21	& 1.16 & 2374.36	& 3.89 & Yes \\
		\hline
	\end{tabular}
	\caption{Pov-ray experiment. Comparison between the execution time (s) for Scenarios 1 ($\mu_1$) and 2 ($\mu_2$) and hypothesis testing.} 	\label{tab:pov-ray}
\end{table}
%\normalsize

\subsection{I/O intensive application}
We used IOzone as a representative benchmark of an I/O application. Table~\ref{tab:iozoneExperiment} depicts the results (in seconds) for the execution of the iozone benchmark -- the benchmark was executed as follows:  \textit{iozone -a -i 0 -i 1 -g 4M}.  If we compare both scenarios, we can conclude that there is enough statistical evidence to accept $H_0$:  As the number of pods in a machine increases, the caused overhead is higher.  The conclusion of these experiments is that it is better to group all the containers in the same pod.
\begin{table}[t]
\footnotesize
	\renewcommand{\arraystretch}{0.9}
	\centering
	\begin{tabular}{| c | c | c | c | c|c|}
	%	\hline
		\multicolumn{1}{c}{}& \multicolumn{1}{c}{\textbf{Scenario 1}} & \multicolumn{1}{c}{} & \multicolumn{1}{c}{\textbf{Scenario 2}} & \multicolumn{1}{c}{} & \multicolumn{1}{c}{} \\
		\hline
		\boldmath$C$ & \boldmath$\mu_1$ & \boldmath$\sigma_1$ & \boldmath$\mu_2$ & \boldmath$\sigma_2$ & \boldmath$\mu_1 - \mu_2 = 0$? \\
		\hline
		1	& 23.52 & 0.82	& 23.19 & 	0.64 & Yes \\
	    4	& 60.85 &	1.45  &	65.02	& 1.25 & No \\
	    8   &  85.98  & 2.25  & 91.36 & 2.24 & No \\
		12	& 108.54 &	4.14	 &	91.36	& 3.40 & No \\
		20	& 153.51 &	6.47 	& 170.99 	& 5.28 & No \\
		\hline
	\end{tabular}
		\caption{IOzone experiment (iozone -a -i 0 -i 1 -g 4M). Comparison between the execution time (s) for Scenario 1 ($\mu_1$) and for  Scenario 2 ($\mu_2$) and hypothesis testing.}	\label{tab:iozoneExperiment}
\end{table}

\subsection{Network intensive application}
%\subsubsection{Network intensive application}
%\footnotesize
\begin{table}[t]
\footnotesize
	\renewcommand{\arraystretch}{0.9}

	\centering
	\begin{tabular}{|c|c|c|c|c|c|c|c|}
		\multicolumn{1}{c}{}& \multicolumn{1}{c}{\textbf{Scenario 1}} & \multicolumn{1}{c}{} & \multicolumn{1}{c}{} & \multicolumn{1}{c}{\textbf{Scenario 2}}& \multicolumn{1}{c}{} & \multicolumn{1}{c}{} & \multicolumn{1}{c}{}\\
		\hline
		\boldmath$C$ & \boldmath$\mu_1$ & \boldmath$\sigma_1$ & \boldmath$\sum \frac{BW_i}{C}$ & \boldmath$\mu_2$ & \boldmath$\sigma_2$ & \boldmath$\sum \frac{BW_i}{C} $ & \boldmath$H_0?$  \\
		\hline
	1 &	1.88	& 0.06	& 1.88 & 	1.90	& 0.04	& 1.90 &	Yes\\
	4 &	8.61	& 0.21	& 2.15 & 	8.82	& 0.05	& 2.20 &	Yes \\
	8 & 	15.53	& 0.12	& 1.94 & 16.26	& 0.20	& 2.03 &	 No \\
	12 &	14.99	& 0.21	&  1.25 & 16.42 &0.38	& 1.37 &	 No \\
	20 &	15.10	& 0.19	& 0.75 &	18.32	& 0.91	& 0.91 &	No \\
%	40 & 	17.33 &	0.25	& 0.43 &	25.21	& 1.75	& 0.63 &	No  \\	
		\hline
	\end{tabular}
		\caption{Hypothesis test for Network bandwidth (GB) for $C$ iperf Clients. Iperf server \& client are on the same machine.}
	\label{tab:iperf1}
\end{table}
%\normalsize

%\small
\begin{table}[t]
\footnotesize
	\renewcommand{\arraystretch}{0.9}

	\centering
	\begin{tabular}{|c| c| c | c | c | c | c|c|}
		%	\hline
		\multicolumn{1}{c}{}& \multicolumn{1}{c}{\textbf{Scenario 1}} & \multicolumn{1}{c}{} & \multicolumn{1}{c}{} & \multicolumn{1}{c}{\textbf{Scenario 2}} & \multicolumn{1}{c}{} & \multicolumn{1}{c}{} & \multicolumn{1}{c}{}\\
		\hline
		\boldmath$C$ & \boldmath$\mu_1$ & \boldmath$\sigma_1$ & \boldmath$\sum \frac{BW_i}{C}$ & \boldmath$\mu_2$ & \boldmath$\sigma_2$ & \boldmath$\sum \frac{BW_i}{C} $ & \boldmath$H_0$? \\
		\hline
		
	1 &	108.26 &	0.04 &	108.26 &	108.26 & 0.04 &	108.26 & Yes\\
	4 &	110.53 &	0.39 &	27.63 &	110.17 &	0.84 &	27.54 & Yes \\
	8 &	113.81 &	0.47 &	14.23 &	115.64 & 	0.51 & 	14.46 & No\\
	12 &	117.42 &	0.92 &	9.78 &	117.53 &	4.01 &	9.79 & Yes \\
	20 &	124.74 & 	1.22 &	6.24 &	126.52 &	2.09 & 6.33 &  No \\
		\hline
	\end{tabular}
		\caption{Hypothesis test for Network bandwidth (MB) for $C$ iperf Clients.  Iperf Clients are on a different physical machine from the server one.}
	\label{tab:iperf2}
\end{table}

The network infrastructure of a machine is shared by all the containers inside a pod. All the containers in a pod share the port space and the pod has only one IP address.
Sharing the access to the network by several containers might cause an overhead on the container performance.  
To measure that overhead, we deployed an {\tt iperf} server inside a pod and several clients with the previous scenario configuration.  All test measure the network bandwidth for a 30 second interval of TCP traffic.

The first experiment schedules all the containers at the same machine.  In a real scenario, this situation might arise when the scheduler groups pods / containers together with high network traffic among them. In Table~\ref{tab:iperf1}, we show the average bandwidth per container and the hypothesis test.  When the number of containers inside the pod is above 4, there is enough statistical evidence to reject $H_0$.  The best results are achieved when each pod has an isolate container (Scenario 2).  

We repeated the experiments with the {\tt iperf} server placed on a machine and the clients scheduled in another machine. Table~\ref{tab:iperf2} shows the results, which are similar to the previous ones.  The bandwidth values from Scenario 2 are higher than the values from Scenario 1. From these experiments, we can conclude that deploying several pods with a few coupled containers is better than a single pod with a large number of containers.

\section{Discussion}
\label{sec:discus}
% ???
%\subsection{Rules to structure the deployment of an application}
% Victor

\begin{figure}[t]
	\centering
	\includegraphics[width=0.9\linewidth]{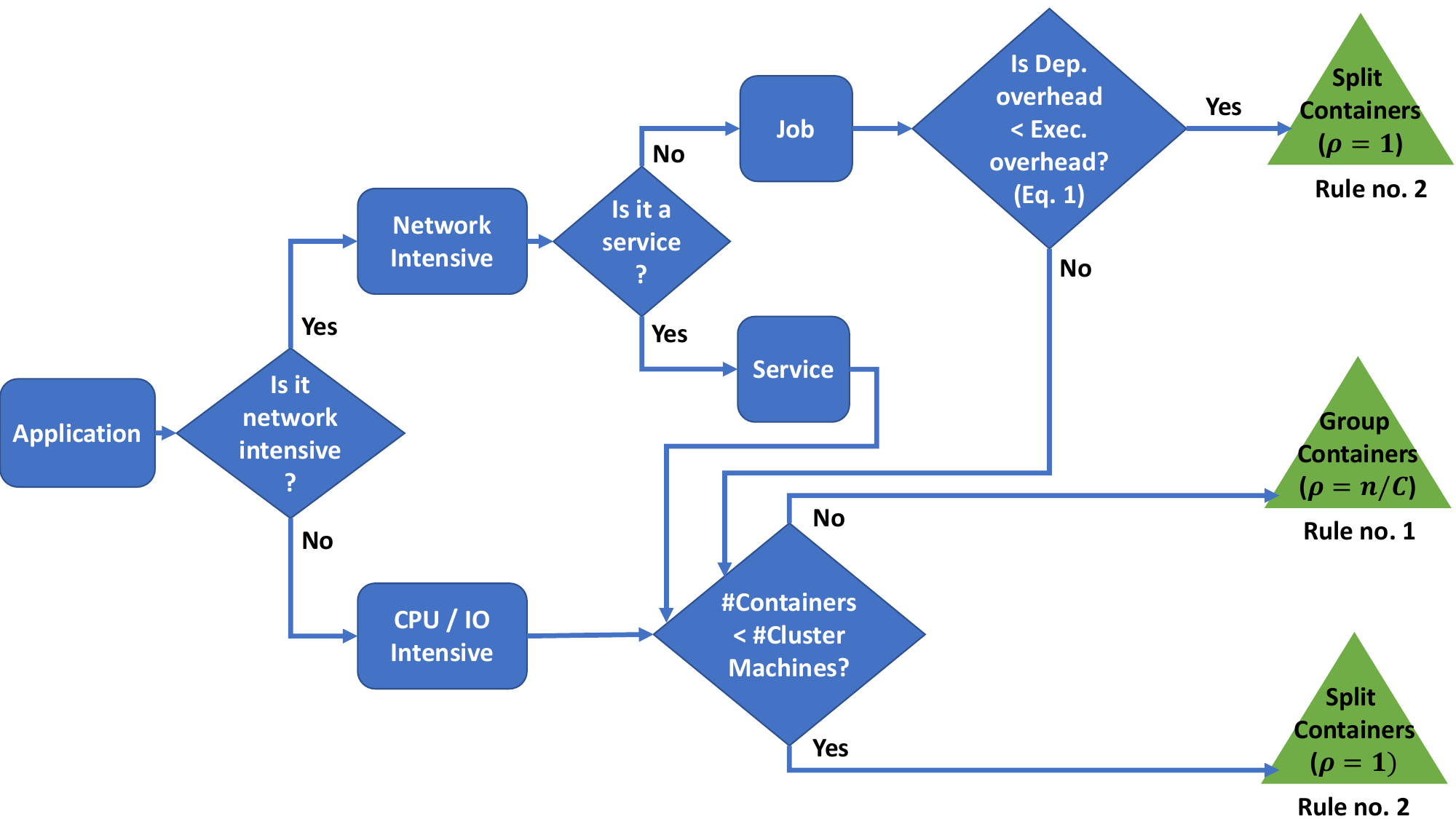}
	\caption{Flow diagram to choose the best $\rho$ parameter. $C$ is the number of containers to deploy and $n$ is the number of machines in the cluster.}
	\label{fig:tree}
\end{figure}

We demonstrated that the deployment of an application on a specific infrastructure can impact its overall  performance.  We used results from our experiments to derive rules that try to improve it.  Since in Kubernetes the minimal schedulable unit is the pod, then $\rho$ is the parameter which has the highest impact on performance.   We assume that the Kubernetes nodes are homogeneous and that all the containers can be distributed across physical nodes, improving the performance of the application -- i.e. there is no coupling between containers, and this is considered as a design restriction.  Figure~\ref{fig:tree} summarises the rules to choose the best $\rho$ from our experiments.

%\begin{table*}[t!]
%\center
%\footnotesize
%	\renewcommand{\arraystretch}{1}
%
%    \begin{tabular}{ | c | l | c |}
%     \hline
%      \textbf{Rule no. }& \textbf{Condition} & \boldmath$\rho$ \\ \hline %\hline
%   %  \multicolumn{3}{|l|}{\textbf{Rules for Task/job Pods}} \\
%   %  \hline
%      1 &   \splitbox{Task/job Pod \&  CPU intensive app.} & $\rho = \frac{n}{c}$   \\
%      \hline
%      2 &  \splitbox{Task/job Pod \& Net. intensive app.   \&  Eq. \ref{eq:1}} & $\rho = 1$  \\ 
%      \hline
%   % \multicolumn{3}{|l|}{\textbf{Rules for Service Pods (\boldmath$T_d$ negligible)}} \\
%   %  \hline
%     3 &   \splitbox{Service Pod \& CPU intensive app.}  & $\rho = \frac{n}{c}$    \\
%     \hline
%     4 &  \splitbox{Service Pod \& Net. intensive app.} & $\rho= 1$  \\
%     \hline
%   %  \multicolumn{3}{|l|}{\textbf{Default rules}} \\
%  %   \hline
%     5 & $c \ge n$ & $\rho =  \frac{n}{c}$ \\
%     \hline
%     6 & $c < n$ & $\rho =  1$ \\
%     \hline
%    \end{tabular}
%
%
%\caption{Rules to choose the best $\rho$ parameter. $c$ is the number of containers to deploy and $n$ is the number of machines in the cluster. Rules with patterns more specific have high priority. Rules 5 \& 6 are the default rules.}
%\label{tab:rules}
%\end{table*}

If an application is CPU or I/O intensive, it is better to group all the containers together --from experiments in Tables~\ref{tab:pov-ray} and \ref{tab:iozoneExperiment}.  However,  we want to distribute the pods among as many machines as possible --through $\rho$ parameter. If the number of containers is greater than the number of machines, $\rho$ should be equal to $\frac{n}{c}$ (\textbf{Rule no. 1}) -- we have $n$ pods with $\frac{c}{n}$ containers at each pod.  This rule tries to minimise the impact of $T_d$ -- which decreases for low values of $\rho$.  If the number of containers to deploy is less than the number of machines, then $\rho=1$ (\textbf{Rule no. 2.}) -- we deploy a pod with a container at each machine. 

\begin{figure}[t]
	\centering
	\includegraphics[width=0.5\linewidth]{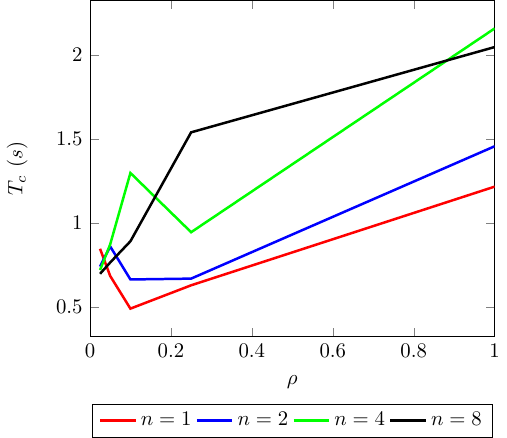}
	\caption{Function $T_t(\rho)$ for different values of $n$. The number of deployed containers is assumed to tend to infinity}
	\label{fig:Example}
\end{figure}

The $\rho$ choice is different if we consider an application that makes a high use of the network.  If it is a service pod and there are few failures in the scenario --equivalently, $T_d$ is negligible -- the best choice is to set $\rho = 1$. The reason is that regardless of the machine where a pod is scheduled, the effective bandwidth is higher when there is only one container inside a pod (Tables~\ref{tab:iperf1} and \ref{tab:iperf2}).  However, if $T_d$ is relevant, we can calculate the total time $T_t$ (deployment time $T_d$ plus execution time $T_e$) as a function of $\rho$: $T_t(\rho) = T_d(\rho) + \alpha(\rho)T_e = \frac{c}{n}T_c(\rho)+ \alpha(\rho)T_e$; where $\alpha(\rho)$ is the overhead caused by the pod abstraction (Section~\ref{sec:experiment}) and it can be calculated as $\frac{\mu_1}{\mu_2}$, where $\mu_2$ corresponds to a scenario with $\frac{1}{\rho}$ containers per pod. In general, as $\mu_2$ is expected to be greater than $\mu_1$, then $\alpha >1$. Additionally, $\alpha(1) = 1$.  Figure~\ref{fig:Example} depicts an example of $T_c(\rho)$, calculated when $C\to\infty$, obtained from Figure~\ref{graph:deployment2}.

It is a complex task to minimise the function $T_t(\rho)$.  As a simplification, we can assume that $\alpha(\rho)$ remains constant and when $n$ tends to infinity, the mean time to create a container also remains constant.  In our experiments, for low values of $\rho$ (Table \ref{tab:iperf2}), its value is about 1.01.  Assuming that the major improvement in the execution time is achieved by executing tasks in parallel, we can compare the situation where $\rho = 1$ versus $\rho=\frac{n}{c}$.  The first one will be faster than the second one when Equation~\ref{eq:1} is satisfied -- \textbf{Rule no. 2} should be applied.  Otherwise,  \textbf{Rule no. 1} will be more suitable.
$$T_t(1) < T_t(\sfrac{n}{c}) \implies \frac{c}{n}T_c(1) + T_e < \frac{c}{n}T_c(\sfrac{n}{c}) + \alpha T_e \implies$$ 
\begin{equation}\label{eq:1}
T_c(1) - T_c(\sfrac{n}{c}) < \frac{n(\alpha-1)}{c}T_e
\end{equation}

%If Equation~\ref{eq:1} is satisfied, then the \textbf{Rule no. 2} is applied.  Otherwise, the \textbf{Rule no. 1} will be more suitable. % Finally, as our experiments show that for intensive I/O applications the value of $\rho$ has not influence on the performance (Table~\ref{tab:bzipExperiment}); for that applications the default rule is applied.  Otherwise, if there are less containers than the number of machines ($c < n$), then we should split the containers across as machines as possible ($\rho= 1$) (\textbf{Rule no. 6}).

These rules are based on the experiments in Section~\ref{sec:experiment}.  Other container technologies -- such as Linux LXC or Core OS rocket -- can be abstracted in a similar way.  The use of a particular technology does not have an impact on our model, as many of these container framework will also share similar lifecycle states.  However, the performance values may vary depending on the use of a particular container framework/ technology. In Section~\ref{sec:related}, we provide a comparison of the performance of different technologies. Additionally, there are different container management systems such as Docker Swarm, or Apache Mesos. Although these other platforms do not have the pod abstraction, our models and results could be relevant to them in scenarios where $\rho = 1$.

%As our model is based on Kubernetes and the pod abstraction, several changes should be done in the model to analyse other platforms when $\rho \not= 1$. 

In our work, we have proposed a methodology to feed the model and to analyse the overhead of pod abstraction.  This methodology should be applied to different configurations to be generalised.  For instance, all our experiments were carried out within a private cloud and Kubernetes was deployed over a {\it bare metal} system.  This configuration allows us to avoid the additional overhead caused by the execution of Kubernetes inside VMs.  On the other hand, the Google Cloud Engine platform gives the possibility of running a Kubernetes cluster; however, the containers are run over VMs, which may have an impact on the performance and the hypothesis tests may change.  Besides, the underlying service architecture is different.  For example, since the storage service is accesses through the cloud, the I/O intensive application will have a different behaviour, and the overhead caused by pod abstraction may not be negligible.

\section{Related Work}
\label{sec:related}
\begin{table}
		\centering
	\renewcommand{\arraystretch}{0.9}
	        \footnotesize
	\begin{tabular}{| c | c | c | c | c|c|}
			\hline
		\textbf{Work} & \textbf{Model} & \splitbox{\textbf{Virtualisation} \\ \textbf{infrastructure}} & \splitbox{\textbf{Experimental} \\ \textbf{framework}}  \\
		\hline
		\cite{hwang2013component}& Experimental approach  & VMs & \splitbox{Hyper-V, KVM, \\ vSphere, Xen} \\ \hline
		\cite{ younge2011analysis} & Experimental approach  & VMs & \splitbox{KVM, Xen, vBox} \\ \hline
		\cite{morabito2015hypervisors, seo2014performance} &  Experimental approach  & \splitbox{Containers \\ and VMs} & \splitbox{Docker, KVM,\\ Xen, LXC}\\ \hline
		\cite{xavier2013performance} & Experimental approach  &  \splitbox{Containers \\ and VMs} & \splitbox{LXC, OpenVZ, \\VServer, Xen} \\ \hline
		\cite{ruiz2015performance} & Experimental approach  & Containers & LXC \\ \hline
		\cite{Amaral:2015,Felter:2015kq} & Experimental approach  & Containers & Docker, KVM\\ \hline
		\cite{kratzke2015performance} & Experimental approach & Containers & Docker, Weave \\ \hline
		\cite{Khazaei:2012rg,khazaei2016efficiency}	& \splitbox{Continuous Markov Chains \\ (Exponential PDF)} & \splitbox{Containers \\ over VMs} & \splitbox{Docker Swarm \\ over Amazon EC2} \\ \hline
		\cite{Merino:2015fk}	& \splitbox{Nets within Nets  (any PDF)} & VMs & Simulations  \\ \hline
		Our work & \splitbox{Nets within Nets  (any PDF) \\ Experimental approach} & Containers & \splitbox{Kubernetes \\over bare metal} \\ \hline

		\hline
	\end{tabular}
	\caption{Summary of related work with the kind of model they proposed, the assumptions of the model, the virtualisation infrastructure that they used and the experimental framework.}	\label{tab:related}
\end{table}

%Performance evaluation is an important  aspect of cloud environments. However, most of the research on this topic has focused on performance comparison of virtual machines rather than containers as the unit of computation. A set of workloads has been developed to get the usage of memory, CPU, networking and storage \cite{Soltesz:2007ys, Felter:2015kq, Amaral:2015}.

We summarised the most important references in performance evaluation of cloud environments in Table \ref{tab:related}. Most of them focused on performance comparison of VMs rather than on containers as the unit of computation. A set of workloads were developed to get the usage of memory, CPU, networking, and storage \cite{Soltesz:2007ys, Felter:2015kq, Amaral:2015}.  However, all of them are based on experimental results that do not have an analytical model that supports reasoning about performance.

To the best of our knowledge, no work has tackled container performance from a rigorous analytical perspective. Even in previous cloud technologies, few studies are based on formal models~\cite{Khazaei:2012rg}.  A Markov Chain based analytical model is used in~\cite{khazaei2016efficiency} to study performance analysis of microservices and implemented with Docker and Docker Swarm.
%providing cluster management support and Amazon EC2 as the virtual machine backend. 
However, their analytical model assumes that the workload generation rate follows a Poisson distribution -- the time between arrivals  has an exponential probability density function (PDF) -- which may yield to non-realistic scenarios. In contrast, we can link Petri Net transitions with any PDF or with functions obtained from real application benchmarking.

% On the other hand, transitions in our Petri Net model can be linked with any probability distribution or with functions obtained from real application benchmarking, so more real scenarios could be modeled. Also, the experimental platform of our proposal is based on Kubernetes.

In~\cite{Merino:2015fk}, an iterative, and step-wise refinement methodology was proposed for cloud applications, covering all the software lifecycle steps. The methodology is centered around a performance model that captures (non) functional requirements, control flow, data flow and the involved computing resources. As a result the performance of the distributed system can be formally analysed by considering all the aspects that affect performance. % Among different modelling flavours,  Timed Petri Nets includes time  to enable exploration of  different performance aspects in a model. Moreover, realistic features could be analysed when PNs are used as a simulation tool. Other paradigms such as queueing networks, cannot take advantage of this possibility. 

All these models need temporal information for feeding them. Previous studies are focused on obtaining this kind of information~\cite{hwang2013component, younge2011analysis}. Some others provide a performance comparison between VMs and containers~\cite{Felter:2015kq, morabito2015hypervisors, seo2014performance}, concluding that there is better performance and resource usage in containers. Scalability performance in Kubernetes was studied in~\cite{joy2015performance}, where the results show that containers perform 22x times faster than VMs for the provisioning action. Container platform evaluation, as Docker and LXC in~\cite{xavier2013performance}, shows performance issues being improved and this approach being considered for High Performance Computing (HPC). They conclude that the performance is near-native for both technologies. In~\cite{ruiz2015performance}, the authors present several variables (e.g. Linux kernel version) having an impact on the performance of containers.  However, both studies focus on the execution time as a performance metric and they do not consider other measures. % They also assume that the image is pre-load in the machines and no methodology was proposed to use their results.  

As we described, Kubernetes introduces the pod abstraction.  To the best of our knowledge, a performance analysis of this abstraction has not been conducted, but the analysis of nested containers is the closest research field.  In~\cite{Amaral:2015, kratzke2015performance}, network performance degradation  was observed in some configurations because of a deployment based on full nested containers.  This degradation is caused by the usage of network virtualization technologies -- Linux Bridge or OpenvSwicht -- twice or by the usage of Software Defined Networks and encryptation.  However, the Kubernetes pod abstraction gives a common space port to all containers -- and therefore the same IP address to all services -- so the performance degradation may be different.
%In this case, network virtualization technologies, as Linux Bridge or OpenvSwitch, are used twice. But, unlike Docker, Kubernetes platform offers the pod concept where the same IP address serves all containers running inside. In this way, better network performance is achieved as network virtualization is used only once.

Finally, several performance metrics of Kubernetes~\footnote{\url{http://blog.kubernetes.io/2016/03/1000-nodes-and-beyond-updates-to-Kubernetes-performance-and-scalability-in-12.html}} were reported by its team. In a cluster of 100 nodes, their results show a 99th percentile pod startup time below 3 seconds, which is consistent with our results.  Their experiments show how a Kubernetes cluster behaves when the scale of the deployment is increased by pod start time, end-to-end response time and response time of different API operations. However, the way the deployment is structured and the impact of grouping containers inside a pod were not analysed.

\section{Conclusions and Future Work}
\label{sec:conc}
An efficacious automated resource management in cloud computing requires to launch, terminate and maintain computing instances rapidly, with a minimum overhead. In this paper, we conduct performance analysis over Kubernetes, achieved through a
Petri Net-based performance model. It allows us to analyse deployment and termination overheads of containers in Kubernetes, as well as understanding the performance of different configurations of a Kubernetes pod -- i.e. the influence of the number of containers per pod. We conducted our analysis in a Kubernetes cluster of 8 machines. %with each machine having 12 Intel Xeon E5-2620 (2.00GHz) cores and 32 GB of RAM.
%Our experiments also considered the impact of downloading the image of a container at deployment time as well as the impact of a heterogeneous latency (Round-Trip Time (RTT)) in the communication network.
Our model can be exploited as a basis to improve two activities: (i) capacity planning and resource management; (ii) application design, specifically how an application may be structured in terms of pods and containers. 
From our experiments, we can see that a single container can be
deployed in a time interval than ranges from less than a second to up to 3 seconds, depending on the circumstances, -- i.e. the number of pods per container, the number of containers deployed simultaneously, the network latency, or the number of host machines. In contrast, the termination time is typically in the order of
a tenth of a second. Moreover, we also provide a set of rules that assist in allocating the number of containers per pod that provides the best performance. These
rules consider a number of characteristics of the application, such as the usage of CPU or network.

As future work, we expect to study the resource contention phenomenon in containers, which appears when multiple containers compete for the same computational resource. We plan to exploit our model with different purposes: (i) to estimate the overhead introduced by resource contention in containers, (ii) to improve the Kubernetes scheduler so that it is aware of resource contention problems and (iii) to undertake various {\it what-if} scenarios to investigate the behaviour of different resource management policies. 
%if an application has $c$ containers, the number of available host machines in the cluster is $n$, with $c\gtn$, and the images of the containers are preloaded at each machine, then, in most of the cases, the number of containers per pod that provides the best performance is given by $n/c$.  This default rule tries to split the parallelism of the cluster and it minimise the deployment time.  However, if there is a high usage of the network it should be better to put one container per pod.

%Once the model has been built and tuned with micro-benchmark results, we can use it with different purposes: i) to estimate the overhead introduced by interference.  It allows developer to test different application deployment configurations.  ii)  The Scheduler can take into account  applications  sensitivity to interference  to allocate containers per node.  iii) The model can be used to undertake various {\it what-if} scenarios to investigate the behaviour of different resource management policies.  
%Analysis of formal model and experiments.

%Simulations of complex scenarios

%Metrics about the life-cycle

%Monitoring and controlling a Kubernetes Cluster

%\input{Files/Conclusion.tex}

\section*{Acknowledgments}
This work was supported in part by: The Industry and Innovation department of the Aragonese Government and European Social Funds (COSMOS group, ref. T93) and the Spanish Ministry of Economy (“Programa de I+D+i Estatal de Investigaci\'{o}n, Desarrollo e innovaci\'{o}n Orientada a los Retos de la Sociedad” –TIN2013-40809-R). V.~Medel was the recipient of a fellowship from the Spanish Ministry of Economy.

% and from the Fundaci\'{o}n Ibercaja-CAI.

\bibliographystyle{IEEEtran}

\end{document}